\documentclass[10pt,conference]{IEEEtran}
\IEEEoverridecommandlockouts
\usepackage[T1]{fontenc}
\usepackage{amsmath,amssymb,amsfonts}
\usepackage{graphicx}
\usepackage{textcomp}

\usepackage{bbding}%for symbols

\newcommand\YES{\scriptsize\Checkmark}
\newcommand\YEA{\textcolor{gray}{\scriptsize\Checkmark}}

\usepackage[inline]{enumitem}
\usepackage{placeins}
\usepackage{balance}

\usepackage[framemethod=tikz]{mdframed}
\mdfdefinestyle{mpdframe}{
    frametitlebackgroundcolor   =black!15,
    frametitlerule              =true,
    roundcorner                 =3pt,
    middlelinewidth             =0.8pt,
    innermargin                 =0.1cm,
    outermargin                 =0.1cm,
    innerleftmargin             =0.1cm,
    innerrightmargin            =0.1cm,
    innertopmargin              =0.1cm,
    innerbottommargin           =0.1cm
}

\usepackage{caption,subcaption}
\usepackage{pgfplots,pgfplotstable}
\pgfplotsset{compat=1.14} 
\usetikzlibrary{pgfplots.groupplots}
\usetikzlibrary{external}
\usepackage{fontawesome}

\usepackage{array,makecell,tabularx}
\usepackage{multirow}
\usepackage{longtable}
\usepackage{booktabs}
\usepackage{rotating}

\usepackage{xspace}
\usepackage[autolanguage]{numprint}
\AtBeginDocument{\npthousandsep{\,}}
\usepackage{fp}

\newcommand*\np[2][z]{%\textcolor{red}{%
\ifx z#1%
$\numprint{#2}$%
\else%
$\numprint[#1]{#2}$%
\fi\xspace%
}

\newcommand{\ShowAbsoluteNumber}[1]{%
\ifnum #1<10%
{\hspace*{0pt}#1}%
\else%
\ifnum #1<100%
{\hspace*{0pt}#1}%
\else%
\ifnum #1<1000%
{\hspace*{0pt}#1}%
\else%
{\numprint{#1}}%
\fi%
\fi%
\fi%
}

\newcommand{\ShowPercentage}[2]{%
\FPeval\percentage{round(#1/#2*100,0)}%
\FPeval\percentageOneDecimal{round(#1/#2*100,1)}%
\ifnum \percentage=0%
{\np[\%]{\FPprint{percentageOneDecimal}}}%
\else%
\ifnum \percentage<10%
{\np[\%]{\FPprint{percentageOneDecimal}}}%
\else%
{\np[\%]{\FPprint{percentageOneDecimal}}}%
\fi%
\fi%
\xspace
}
\newlength\BARSIZE  
\newcommand{\inlinechart}[2]{%
\FPeval{\BLACKBARSIZE}{#1/#2}\textcolor{black!80}{\rule{\BLACKBARSIZE\BARSIZE}{1.6ex}}%
\FPeval{\BLACKBARSIZE}{1 - (#1/#2)}\textcolor{black!10}{\rule{\BLACKBARSIZE\BARSIZE}{1.6ex}}%
}

\newcommand*\percent[3][v]{%
\ifx b#1% JFF: number/percentage
    (\np{#2} / \ShowPercentage{#2}{#3})\else%
\ifx q#1%
    \np{#2}/\np{#3}(\ShowPercentage{#2}{#3})\else%
\ifx v#1%
    \ShowPercentage{#2}{#3}\else%
\ifx p#1%
    \np{#2}(\ShowPercentage{#2}{#3})\else%
\ifx c#1%
    \inlinechart{#2}{#3}%
\else%
    \np{#2}%
    \ifx r#1%
        /\np{#3}%
    \fi%
    \hspace*{0.5ex}(\ShowPercentage{#2}{#3}) %
    \inlinechart{#2}{#3}%
    \xspace
\fi\fi\fi\fi\fi%
}

\usepackage{solidity}% must be loaded after package "fp", clash of function names

\newcommand\ITEM[1]{\par\smallskip\noindent\textit{\textbf{#1}}}
\newcommand\TODO[1]{%
  \ifx&#1&%
    \textcolor{red}{TODO!} 
  \else
    \begin{flushleft}
      \color{red}%
      TODO: #1%
    \end{flushleft}
  \fi
  \typeout{*** TODO: #1}%
}

\AtBeginDocument{%
}

\usepackage{siunitx}

\catcode`_=\active
\catcode`_=8

\newcommand\SB[1]{\texttt{#1}}
\newcommand\TAG[2][\relax]{%
  \medskip\par\noindent\SB{#2}%
  \ifx\relax#1\else
    \ (#1)%
  \fi
  .
}

\ExplSyntaxOn
\cs_set_eq:NN \explower \text_lowercase:n
\ExplSyntaxOff

\usepackage{swc}% swc needs hyperref

\newcommand\SMARTBUGS{SmartBugs\xspace}
\newcommand\SARIF{SARIF\xspace}

\definecolor{one}{HTML}{4F81BD}
\definecolor{two}{HTML}{C0504D}
\definecolor{three}{HTML}{9BBB59}
\definecolor{more}{HTML}{9F4C7C}

\definecolor{mythril}{HTML}{D6D2D2}
\definecolor{conkas}{HTML}{F1E4F3}
\definecolor{osiris}{HTML}{F4BBD3}
\definecolor{madmax}{HTML}{F686BD}
\definecolor{ethainter}{HTML}{FE5D9F}
\definecolor{securify}{HTML}{9B7874}
\definecolor{teether}{HTML}{666370}
\definecolor{pakala}{HTML}{1C1F33}
\definecolor{ethor}{HTML}{A6B1E1}
\definecolor{maian}{HTML}{B4869F}

\definecolor{bad}{HTML}{FAAAAD}
\definecolor{mid_bad}{HTML}{FCE6E9}
\definecolor{mid_good}{HTML}{BDE2C8}
\definecolor{good}{HTML}{63BE7B}

\definecolor{midblue}{rgb}{0.70,0.78,0.93}
\definecolor{lightblue}{rgb}{0.85,0.95,0.99}
\definecolor{lightgreen}{rgb}{0.74,0.89,0.78}
\definecolor{lightgray}{HTML}{F3F3F3}
\definecolor{greendark}{HTML}{6EC284}
\definecolor{greenmedium}{HTML}{7AC78E}
\definecolor{greenlight}{HTML}{BCE2C8}
\definecolor{redlight}{HTML}{FCE6E9}
\definecolor{redmedium}{HTML}{FBB6B8}
\definecolor{reddark}{HTML}{F8696B}

\def\nbNewTool{\np{8}}
\def\nbTool{\np{19}}

\title{{SmartBugs 2.0}: An Execution Framework for Weakness Detection in Ethereum Smart Contracts%
\thanks{This project was partially supported by national funds through Funda\c{c}\~ao para a Ci\^encia e a Tecnologia (FCT) under project UIDB/50021/2020. The project was also partially supported by the CASTOR Software Research Centre.}}

\author{%
\IEEEauthorblockN{Monika di Angelo}
\IEEEauthorblockA{\textit{TU Wien}\\
Vienna, Austria \\
{0000-0002-4217-4530}}%
\and
\IEEEauthorblockN{Thomas Durieux}
\IEEEauthorblockA{\textit{TU Delft}\\
Delft, Netherlands \\
{0000-0002-1996-6134}}%
\and
\IEEEauthorblockN{Jo\~{a}o~F.~Ferreira}
\IEEEauthorblockA{\textit{INESC-ID and IST, University of Lisbon}\\
Lisbon, Portugal \\
{0000-0002-6612-9013}}%
\and
\IEEEauthorblockN{Gernot Salzer}
\IEEEauthorblockA{\textit{TU Wien}\\
Vienna, Austria \\
{0000-0002-8950-1551}}%
}
\begin{document}
\maketitle

\begin{abstract}
  Smart contracts are blockchain programs that often handle valuable assets.
  Writing secure smart contracts is far from trivial, and any vulnerability may lead to significant financial losses.
  To support developers in identifying and eliminating vulnerabilities, methods and tools for the automated analysis have been proposed.
  However, the lack of commonly accepted benchmark suites and performance metrics makes it difficult to compare and evaluate such tools.
  Moreover, the tools are heterogeneous in their interfaces and reports as well as their runtime requirements, and installing several tools is time-consuming.

  In this paper, we present SmartBugs 2.0, a modular execution framework.
  It provides a uniform interface to 19 tools aimed at smart contract analysis and accepts both Solidity source code and EVM bytecode as input.
  After describing its architecture, we highlight the features of the framework.
  We evaluate the framework via its reception by the community and illustrate its scalability by describing its role in a study involving 3.25 million analyses.
\end{abstract}


\section{Introduction}

Smart contracts are a fundamental part of blockchain technology, particularly on platforms like Ethereum, where they enable the development of decentralized applications.
Benefits like transparency, trust, and security are paired with potential risks, as malicious actors can exploit vulnerable smart contracts and cause substantial financial losses.
Therefore, there is a pressing need for automated tools that help identify such vulnerabilities.

The goal of this paper is to present \SMARTBUGS 2.0, a modular execution framework that simplifies the execution of analysis tools for smart contracts, facilitates reproducibility, and supports large-scale experimental setups. 
It is open-source and publicly available at \url{https://github.com/smartbugs/smartbugs}.

\ITEM{Methodology.}
\SMARTBUGS supports three modes for analyzing smart contracts: Solidity source code, creation bytecode, and runtime code.
It currently includes \nbTool tools encapsulated in docker images.
With its standardized output format (via scripts that parse and normalize the output of the tools), it facilitates an automated comparison of the findings across tools.
In the context of a bulk analysis, it allows for the parallel, randomized execution of tasks for the optimal use of resources.

\ITEM{Envisioned users.}
\SMARTBUGS is intended for 
\begin{itemize}
\item developers auditing smart contracts before deployment,
\item analysts evaluating already deployed smart contracts,
\item tool developers comparing selected tools, 
\item researchers performing large-scale analyses,
\end{itemize}
and thereby advances the state-of-the-art in the automated analysis of smart contracts.

\ITEM{Engineering challenges and new features.}
Compared to the original version, \SMARTBUGS 2.0 offers the following improvements that overcome several engineering challenges: 
\begin{itemize}
\item support for bytecode as input
\item \nbNewTool additional tools
\item modular integration of new tools
\item support for multiple versions of the same tool
\item generic architecture
\item increased robustness and reliability
\item detection and reporting of tool errors and failures
\item \SARIF as output format
\item mapping of tool findings to the SWC taxonomy\footnote{https://swcregistry.io/}
\end{itemize}
By adding bytecode as accepted input format, the range of smart contracts that can be analyzed by \SMARTBUGS has been extended to programs without source code, including all smart contracts already deployed.
Due to its modular structure, \SMARTBUGS 2.0 can easily be extended with further tools.
The standardized output format and the mapping to a tool-independent taxonomy both facilitate the integration of a comprehensive vulnerability analysis into the development cycle.

\ITEM{Validation studies.}
To showcase the capabilities of \SMARTBUGS 2.0, we present a typical use case that demonstrates how \SMARTBUGS 2.0 has supported the largest experimental setup to date, both in terms of the number of tools and the number of analyzed smart contracts.

\section{Architecture}

\begin{figure*}
  \includegraphics[width=\textwidth]{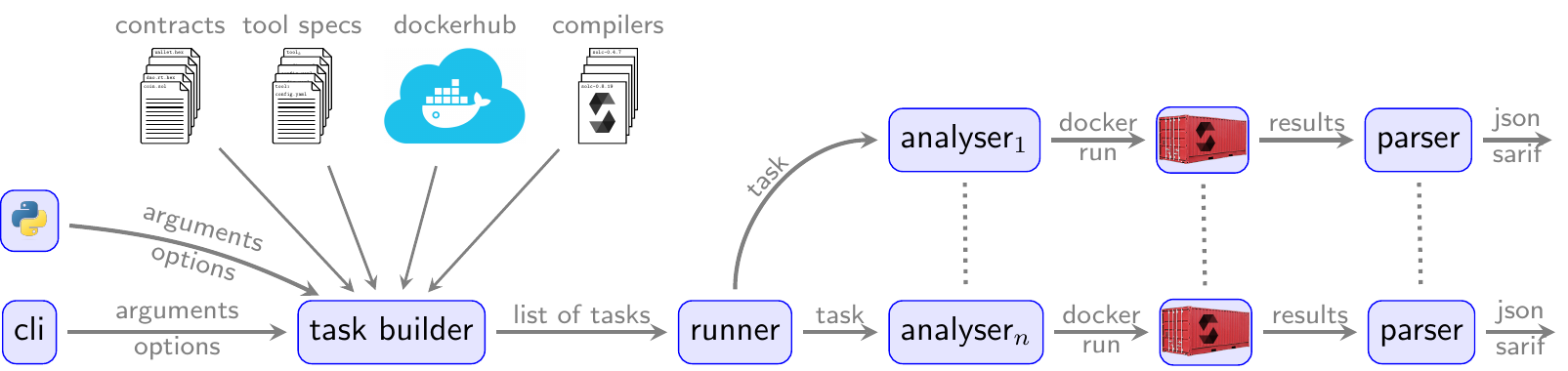}
  \caption{The Architecture of \SMARTBUGS.}\label{fig:arch}
\end{figure*}

Figure~\ref{fig:arch} depicts the architecture of \SMARTBUGS.
It can be started from the command line or called from Python programs.
The main arguments to provide are a specification of the smart contracts to process and a list of tools to execute.
For a mass analysis, it is also important to specify the number of parallel processes as well as resource bounds per process.

\ITEM{Task builder.} For each smart contract matching the specification, the task builder selects those tools that fit the format of the smart contract (source code, creation bytecode, or runtime code) and pulls their Docker images. % from docker.com. %JFF Note: I left this out because 1) it should be Docker Hub; 2) if we do docker login to a different repo, it will probably use that repo?
Moreover, it determines a unique folder for the output of each run.
Sometimes the naming scheme specified by the user leads to collisions, meaning that the output of different smart contracts or tools would end up in the same folder.
The task builder resolves conflicts in a deterministic way such that any restart of \SMARTBUGS with the same arguments after an interrupt leads to the same output folders.

Most tools analyzing Solidity source code either contain a compiler for a fixed Solidity version or download an appropriate compiler on the fly.
Both approaches are problematic in the context of a bulk analysis.
In the first case, the integrated compiler is not able to handle smart contracts written for a different version, whereas in the second case an adequate compiler will be downloaded, but used only once and then discarded together with the container of the tool, which leads to redundant downloads during the analysis.
Therefore, the task builder inspects the smart contracts and downloads the corresponding compilers beforehand.
Later on, during analysis, a compiler matching the smart contract is injected into the container such that the tool is able to compile the contract without attempting to download the compiler itself.
Overall, the task builder downloads all resources and detects problems before actually starting the analysis.
This prevents racing conditions, errors popping up only during the analysis phase, and minimizes network traffic.

\ITEM{Runner.} The runner receives a list of tasks, where each task contains the information for applying a single tool to a single smart contract.
The length of the list is roughly the product of the number of smart contracts and the number of tools.
To improve the utilization of server resources, the runner randomly permutes the task list.
Then it starts the requested number of parallel analyzers, which process the tasks from the list one after the other.

\ITEM{Analyzers.}
Each analyzer picks a task from the queue of the runner, copies the smart contract, the Solidity compiler (if necessary) and auxiliary scripts to a temporary volume and runs the Docker image of the tool with this volume mounted.
Once the Docker container has terminated, the analyzer extracts the result files and writes them to the designated output folder.
It adds a file with meta information like the execution time, the arguments of the Docker run, and the version of the tool.

\ITEM{Parsing.}
The output of the tools is heterogeneous: some provide their results in structured form, others produce textual output.
Parsers are small scripts accompanying each tool. They scan the results for the weaknesses detected, but also watch out for errors (irregular conditions reported by the tool) and failures (exceptions not caught by the tool).
The information is written to JSON files and\,---\,to facilitate the integration of \SMARTBUGS\ into CI workflows\,---\,to \SARIF files. 

\section{Features}

\ITEM{Output format \SARIF.}
\SMARTBUGS 2.0 can provide the results in \SARIF (Static Analysis Results Interchange Format), an OASIS standard that defines a common reporting format for static analysis tools~\cite{SARIF}.
\SARIF is JSON-based and allows IDEs to access the analysis reports in a uniform way.
By adopting a common format that can be parsed by readily available tools, the cost and complexity of aggregating the results of analysis tools into common workflows diminishes.
For example, it becomes trivial to integrate \SMARTBUGS into GitHub workflows, since GitHub automatically creates code scanning alerts in a repository using information from \SARIF files.%
\footnote{\url{https://docs.github.com/en/code-security/code-scanning/integrating-with-code-scanning/uploading-a-sarif-file-to-github}}
%\footnote{\url{https://docs.github.com/en/code-security/code-scanning/integrating-with-code-scanning/sarif-support-for-code-scanning}}
For an example of the integration of \SARIF produced by \SMARTBUGS and GitHub, we refer the reader to the repository \texttt{smartbugs/sarif-tests}.\footnote{\url{https://github.com/smartbugs/sarif-tests/security/code-scanning}}

\ITEM{Bytecode input.}
On Ethereum, smart contracts are deployed by sending a transaction containing the \emph{creation bytecode}.
When executed by the Ethereum Virtual Machine, this code initializes the environment of the new contract and returns the \emph{runtime code} that is actually stored on the chain.
In most cases, the creation bytecode is the result of compiling Solidity source code.
A significant enhancement of \SMARTBUGS 2.0 is its ability to integrate tools that analyze the creation bytecode and runtime code directly, obviating the need to procure Solidity sources first.
In fact, for many smart contracts deployed on the chain, their source code is not available.
Of the \nbTool tools currently included in \SMARTBUGS, \np{13} are able to process creation bytecode and/or runtime code.

\ITEM{Provision of proper compiler versions.}
Another important addition to \SMARTBUGS 2.0 is its ability to select an appropriate compiler for each smart contract.
Solidity has seen a rapid development over the past years, with numerous breaking changes.
Therefore, programmers are strongly advised to include a pragma that specifies the language version that a smart contract was developed for.
Analysis tools have three strategies to cope with this situation.
Experimental tools (proofs-of-concept) may come with just a specific compiler version, restricting its applicability.
Other tools implicitly assume that the compiler on the command search path matches the smart contract to be analyzed.
The most versatile tools inspect the smart contract and download an appropriate compiler before starting analysis.
As none of these approaches fits the needs of an unsupervised bulk analysis, the task builder (see its description above) inspects the smart contracts, downloads each required compiler version once before the actual analysis, and then injects the correct one into every container.
This allows the tool to run the correct compiler version without the need for on-the-fly downloads, which would cost time and increase the network traffic.
As another benefit, this improvement enhances the reproducibility and uniformity of the analyses, as the same compiler version is used consistently across all runs.

\ITEM{Tool integration.}
With the new version of \SMARTBUGS, it is now possible to incorporate new tools without touching the code of \SMARTBUGS.
The details of adding a new tool are described in the wiki of \SMARTBUGS's repository\footnote{\url{https://github.com/smartbugs/smartbugs/wiki/Adding-new-analysis-tools}}.
In essence, a few lines in a configuration file are needed to specify the docker image of the tool and its interface.
Moreover, for extracting the findings and errors from the result files, a Python script has to be added.
This new flexibility in adding tools also allows researchers to compare the behavior of different versions of the same tool, which is particularly useful for evaluating performance over time, or for ensuring that performance does not degrade with an update.

\ITEM{Mapping to a weakness taxonomy.}
To compare and unify findings across tools, the idiosyncratic labels assigned by each tool need to be mapped to a common frame of reference.
\SMARTBUGS 1.0 maps the findings to the vulnerability taxonomy DASP TOP 10\footnote{\url{https://dasp.co/}}.
The new version adds a mapping of all findings (including those of the new tools) to the weakness taxonomy of the SWC registry.\footnote{\url{https://swcregistry.io/}}
The SWC registry is a community-driven catalog of software weaknesses in smart contracts, whose granularity is finer than the one of DASP TOP 10, which allows us to provide more detailed information about the weaknesses found by the tools.
This classification is added to the \SARIF output, in order to be displayed in the context of the source or bytecode.

\ITEM{Supported tools.}
The tools currently in \SMARTBUGS 2.0 are listed in \autoref{tab:tools}. Check marks in black (\Checkmark) indicate new additions, while the gray check marks in column `Solidity' identify the capabilities of the old version.
We added \nbNewTool new tools as well as bytecode support for seven of the old tools.
In most cases, bytecode support refers to runtime code.
Only two tools are able to handle the creation bytecode as well.

\begin{table}[]
  \centering
  \caption{Supported tools.}
  \label{tab:tools}
  \begin{tabular}{@{}llcccc@{}}
  \toprule
	  \multirow{2}{*}{\textbf{Tool}} & \multirow{2}{*}{\textbf{Version}} & \multirow{2}{*}{\textbf{New}} & \multicolumn{3}{c}{\textbf{Contract format}} \\ \cmidrule(l){4-6} 
			&                          & & \textbf{Solidity} & \textbf{Creation} & \textbf{Runtime} \\ \midrule
  \href{https://github.com/christoftorres/ConFuzzius}{ConFuzzius} & \#4315fb7 & \YES & \YES & & \\
  \href{https://github.com/smartbugs/conkas}{Conkas} & \#4e0f256 &  & \YEA & & \YES \\
  \href{https://zenodo.org/record/3760403}{Ethainter} & & \YES & & & \YES \\
  \href{https://secpriv.wien/ethor}{eThor} & 2021 (CCS'20) & \YES & & & \YES \\
  \href{https://github.com/christoftorres/HoneyBadger}{HoneyBadger} & \#ff30c9a &  & \YEA & & \YES \\
  \href{https://github.com/nevillegrech/MadMax}{MadMax} & \#6e9a6e9 & \YES & & & \YES \\
  \href{https://github.com/smartbugs/MAIAN}{Maian} & \#4bab09a &  & \YEA & \YES & \YES \\
  \href{https://github.com/trailofbits/manticore}{Manticore} & 0.3.7 &  & \YEA & & \\
  \href{https://github.com/ConsenSys/mythril}{Mythril} & 0.23.15 &  & \YEA & \YES & \YES \\
  \href{https://github.com/christoftorres/Osiris}{Osiris} & \#d1ecc37 &  & \YEA & & \YES \\
  \href{https://github.com/smartbugs/oyente}{Oyente} & \#480e725 &  & \YEA & & \YES \\
  \href{https://github.com/palkeo/pakala}{Pakala} & \#c84ef38 & \YES & & & \YES \\
  \href{https://github.com/eth-sri/securify}{Securify} & &  & \YEA & & \YES \\
  \href{https://github.com/duytai/sFuzz}{sFuzz} & \#48934c0 & \YES & \YES & \\
  \href{https://github.com/crytic/slither}{Slither} & &  & \YEA & & \\
  \href{https://github.com/smartdec/smartcheck}{Smartcheck} & &  & \YEA & & \\
  \href{https://github.com/protofire/solhint}{Solhint} & 3.3.8 &  & \YEA & & \\
  \href{https://github.com/nescio007/teether}{teEther} & \#04adf56 & \YES & & & \YES \\
  \href{https://github.com/usyd-blockchain/vandal}{Vandal} & \#d2b0043 & \YES & & & \YES \\
  \midrule
  \nbTool tools & & \nbNewTool & 13 & 2 & 13 \\
  \bottomrule
  \end{tabular}
\end{table}

\section{Evaluation}
\ITEM{Reception.}
The appreciation of \SMARTBUGS\ by the community on GitHub is reflected in the following metrics.
With \np{13} contributors, it received over \np{400} stars, \np{81} issues were filed, and \np{110} users/organizations have forked the repository, with 50 unique cloners in the weeks from May 09 to 22, 2023.

\SMARTBUGS\ is not only used by developers and security companies, but also in academic studies \cite{durieux2020, chaliasos2023, qasse2023} or master theses \cite{aryal2021,Veloso-Analise-Estatica-de-Smart-Contracts,Araujo-A-Static-Analysis-based-Platform-as-Service-to-Improve-the-Quality-of-Smart-Contracts,Dinis-Automatic-Bug-Prioritization-of-SmartBugs-Reports-using-Machine-Learning}. Moreover, components of it have been used to build a ML-based tool \cite{mandloi2022}.

\ITEM{Use case.}
In the largest experimental study to date \cite{diangelo2023evolution}, we used \SMARTBUGS 2.0 to execute 13 tools on almost \np{250000} runtime bytecodes. 
The tools reported over \np[million]{1.3} weaknesses in total.
With a resource limit of \np[min]{30} and \np[GB]{32}, the execution took a total of 31 years.
More than half of the tools could run on just \np[GB]{4} for the vast majority of the bytecodes and with less than \np[min]{3} on average per bytecode, while three tools ran into the limits for more than \np{1000} bytecodes.

The new feature in \SMARTBUGS 2.0 of \emph{reporting errors and failures} gives the user an indication, for which bytecodes a tool may be operating outside of its specification. 
This way, potential findings or non-findings are put in relation to the tools ability to properly analyze the bytecode.

\begin{figure}
  \centering
  \includegraphics[width=\linewidth]{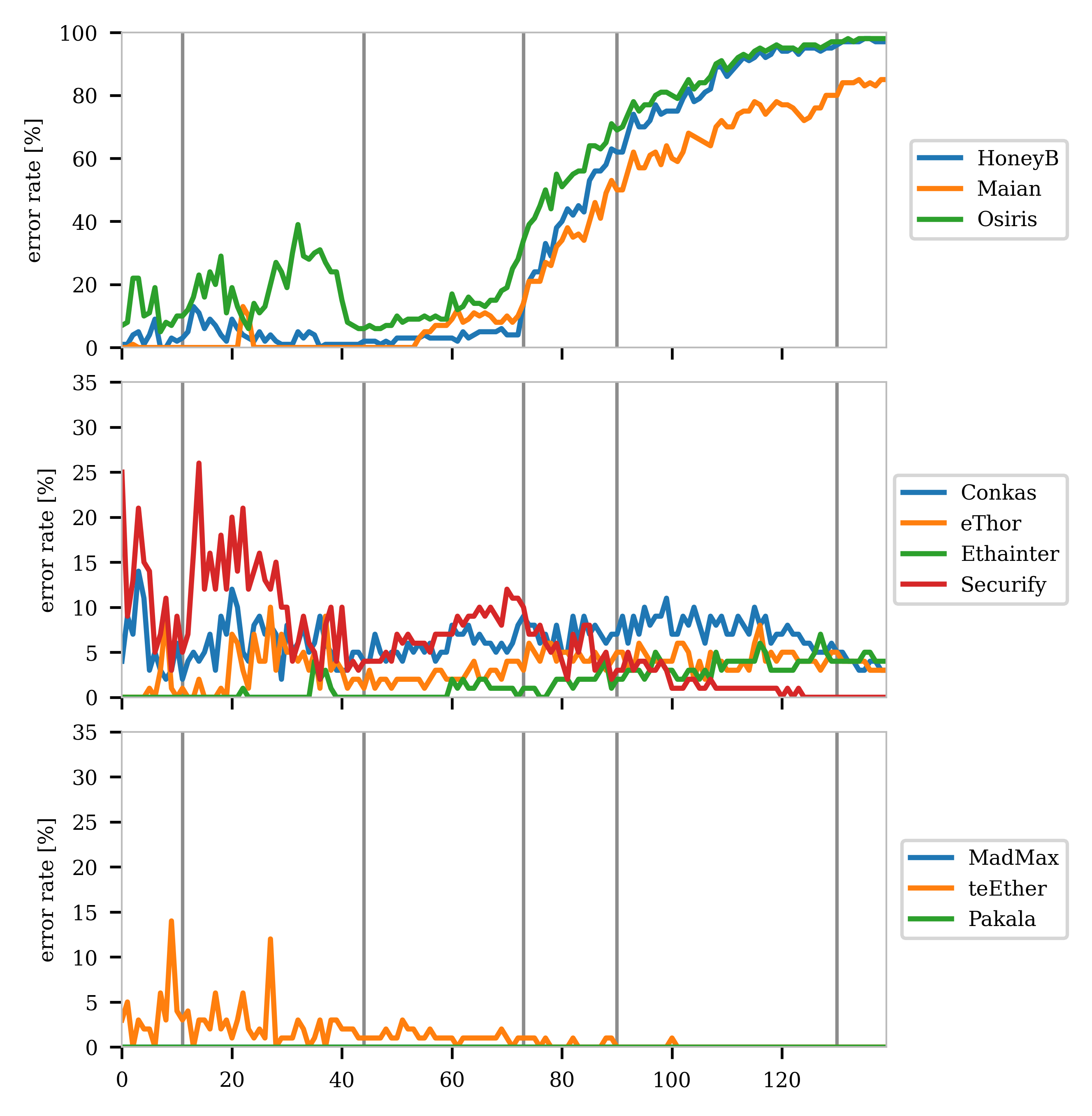}
  \caption{Tool errors over time: percentage of errors encountered by the tools, in bins of \np{100000} blocks.}
  \label{fig:errors}
\end{figure}

\autoref{fig:errors} depicts the error rate of each tool on a time line of blocks on the Ethereum main chain, where each data point represents the percentage of reported errors in bins of \np{100000} bytecodes.
As Mythril, Oyente and Vandal report no errors, they are not depicted.
Apparently, HoneyBadger, Maian, and Osiris experience an increasing error rate after \np{7.5} million blocks.
This information can be used to enhance the tools or make informed decisions about whether to use them for more recent smart contracts.

Moreover, tool failures may serve as a measure of robustness.
For eight tools, the failure rate was below \np[\%]{1} of the bytecodes, whereas for one tool, the failure rate reached \np[\%]{25}, meaning that the tool ran into an exception for one out of four bytecodes.

\section{Related work} 

As documented in the previous sections, \SMARTBUGS 2.0 is a major improvement over the original version of \SMARTBUGS~\cite{ferreira2020smartbugs}, which was released in 2019.
To the best of our knowledge, the only other execution framework that implements similar ideas is USCV~\cite{Ji2021}.
It comprises eight tools for the analysis of Solidity source code, with seven of them also covered by \SMARTBUGS.
USCV seems to be neither widely used nor maintained, as the latest of its 10 commits is from mid-2021 and no issues have been filed so far.
% No citation of ESAF by LOPEZVIVAR doi.org/10.1016/j.comcom.2021.03.008 since it is not open source

\section{Conclusion}

\SMARTBUGS 2.0 has proven to be a useful tool for our own work as well as for fellow researchers and developers.
Its extensive use has shown some limitations, partly resulting in enhancement requests by users.
In future work, we will consider the following extensions.

\emph{Support for historic compiler versions.}
\SMARTBUGS supports Solidity 0.4.11 and above.
By accessing another repository, we can include versions down to 0.4.0.
Compiler versions older than that may be harder to come by.

\emph{Support for more complex formats of source code.}
At the moment, each smart contract has to be contained in a single file.
However, complex projects are split into several files, with additional includes of system-wide libraries.
\SMARTBUGS could try to determine the dependencies and transfer them also into the container.

\emph{Use of source code mappings.}
Tools for bytecode analysis can be made to analyze source code by compiling the source code before feeding the result to the tool. The difficult part is to map the bytecode addresses of weaknesses back to source code lines.

\emph{Addition of new tools.}
The automated analysis of smart contracts is an active area, with new tools emerging every year.
We hope that we will be able to keep up, not least with the help of the community contributing further tool configurations.

% \emph{Refurbishing unmaintained tools.}
% For some tools in \SMARTBUGS that have been frequently cited as a reference it may be worthwhile to keep them up-to-date.
% This includes upgrading the tool to new versions of dependencies, but also extending it to cope with new EVM instructions introduced at forks after the tool was published.

\balance
\bibliographystyle{IEEEtran}
\bibliography{references}
\balance

\end{document}